\documentclass[doc,12pt,natbib]{apa6}
\usepackage{caption}
\usepackage{graphicx}
\usepackage{amsmath} 
\usepackage{amssymb}
\usepackage{amsthm}
\usepackage{mathtools}
\usepackage{array}
\usepackage{url}
\usepackage{bm}
\usepackage{subcaption}

\usepackage[doublespacing]{setspace}
\usepackage[ruled,vlined]{algorithm2e}
\theoremstyle{definition}

\newcommand{\comm}[1]{}

\title{Algorithms for optimizing model-based incomplete block designs}

\shorttitle{Optimal Exchange Algorithm}
\author{Jonas Bjermo$^{1, 2}$, Frank Miller$^{1}$}
\affiliation{$^1$Department of Computer and Information Science, Link\"oping University, Sweden\\
$\mbox{}^2$Department of Statistics, Stockholm University, Sweden}
\date{\today}
\begin{document}

\begin{singlespace}
\maketitle
\begin{center}
\begin{minipage}{\textwidth}
\textbf{Abstract.}
Because of time limitations or participation burden, the treatments in an experimental design can be too large for a single subject. Instead of addressing this using combinatorial incomplete block designs, we propose a model-based approach that optimizes model parameters. This offers distinct advantages: it incorporates subject-specific covariates to tailor treatment allocation to individual characteristics, allows for varying block sizes, and eliminates the equal-treatment replication requirement.
Despite these benefits, model-based approaches are limited by a lack of software and prohibitively large search spaces, making exact optimization computationally intractable. Therefore, we present local search heuristic algorithms and compare them to existing methods.
We evaluate first- and best-improvement algorithms, simulated
annealing (SA), threshold accepting (TA), and two novel algorithms utilizing directional derivatives (dd) to guide exchanges. Serving as discrete versions of continuous gradient-based methods, these dd algorithms take smaller steps and avoid flat regions by prioritizing large-difference dd exchanges.
Our broadly applicable approach uses item calibration in achievement tests as a comparative example to evaluate objective values and computational times. Results demonstrate that for larger problems, the dd algorithms achieve near-optimal solutions significantly faster than SA and TA. Due to computational efficiency, our algorithms offer a highly appealing approach for practical applications.

\end{minipage}
\end{center}
 
\end{singlespace}

\section{Introduction}\label{introduction}
Designing efficient experiments is of great interest across many application areas.
A common situation arises when several treatments need to be investigated, and multiple subjects are recruited to test them. Often, each subject can receive several treatments, but the total number of treatments is too large for any single subject to evaluate all of them, either because of time limitations or concerns about participant burden.

To address this, researchers use incomplete block designs, where each block corresponds to a subject, and each subject is assigned only a subset of the treatments \citep[see, e.g.,][]{dey2010incomplete,nguyen2025incomplete}. The subset size is fixed, and the order of treatments can also be varied if order effects are of concern. 
In medical research, especially in early clinical and preclinical development, such designs are widely applied and are referred to as crossover designs \citep{senn2002cross}. In this setting, incomplete block designs are useful when subjects can only receive a limited number of treatments \citep[Chapter 7.3 of][]{senn2002cross}. 
Another important application is in sensory testing \citep[Chapter 4.8]{stone2020sensory}, where participants may be asked to evaluate many products, but practical limitations such as fatigue restrict the number of products each subject can test. 

Incomplete block designs are created traditionally in a combinatorial way, ensuring balance and symmetry in the use of treatments in blocks. While we consider here the same setup with treatments allocated to subjects, we focus, in contrast, on model-based designs assuming parametric models for each treatment. Based on the models, we can quantify the information about the parameters that we obtain in the experiment, given a specific design. The selection of a design aims to optimize the parameters' information. Using this model-based approach to choose optimal designs rather than the combinatorial approach has a couple of advantages: 
(1) It is easy to handle subject-specific covariates in the models, which enables us to tailor the treatment allocation to the individual characteristics;
(2) we are not bound to use the same block size (number of treatments that a subject receives), which can be good in situations where treatments have different burdens or time requirements on individuals;
(3) it is not required to have an equal number of replicates of treatments in the study. 
On the other hand, a limitation of the model-based approach is the dependence on these models: It is crucial that the models chosen have their justification in the application area.

Another disadvantage of the model-based approach, in contrast to the combinatorial-based incomplete block designs, is that software to determine optimal designs is less developed. For combinatorial-based designs, construction methods are, e.g., implemented in several {\tt R}-packages \citep{R-blocksdesign, R-crossdes, R-ibd}, and it is also possible to use {\tt proc optex} in the SAS software \citep{SalinasRuiz2024}. This article aims to develop and improve algorithms for optimizing model-based incomplete block designs.  

For example, consider a sensory testing experiment in which each subject can only receive a subset of the available treatments (food products to taste) due to time and risk of subject fatigue. Suppose there are 20 treatments, but each individual can be assigned to no more than 5 of them. When we have, say, 40 individuals, the number of possible treatment allocations gets huge: Each individual could receive ${ 20 \choose 5 }=15504$ different subsets of treatments and for the 40 subjects, we have ${ 20 \choose 5 }^{40}\approx 4.1\cdot 10^{167}$ different designs, which means that exact optimization methods searching the optimal allocation out of all possible ones will be unfeasible. 

Heuristic methods based on local search techniques are very common to use in these kinds of combinatorial problems. Local search algorithms start from a random solution and then select a new better solution from a neighborhood of the previous solution. The search continues until no further improvements are possible. The neighborhood search can, e.g., be done by a first or best improvement procedure.

In the experimental optimal design literature, these methods are called the best improvement exchange algorithms \citep{Fedorov} and the first improvement exchange algorithm \citep{CookNachtsheim1980}. However, they often get trapped in local optima too far from the global optimum. It is, therefore, common to implement "up-hill" moves in the local search algorithm to prevent it from getting trapped in local optima with objective function values too far from the global one.

One of the most common algorithms is simulated annealing \citep{KirkpatrickGelattVecchi}, 
which has nice asymptotic properties if a large number of iterations is allowed, and it achieves an approximate solution close to the global optimum.

Since simulated annealing is slow, we also explore other possible faster algorithms with equally or better final results as simulated annealing. \cite{DUECK1990161} presented the threshold accepting method, which has a simpler structure than simulated annealing and is superior both in speed and final objective function values for several examples. We will construct a threshold accepting method for our purpose. 

Since we adopt a model-based approach, we suggest using directional derivatives for wiser choices of exchanges in the neighborhood to speed up the algorithms. We have developed a first improvement algorithm using directional derivatives that serves as a very fast alternative to the already established algorithms.

We also suggest using the iterated local search algorithm \citep{Talbi}, combined with threshold accepting, to find approximately good solutions relatively fast. 
This is similar to the chained local optimization algorithm described in \cite{Martin1996}, but it used simulated annealing.

An example of an experimental optimal design problem that is suitable for the model-based optimization algorithm is item calibration in achievement tests \citep{Miller_Fackle-Fornius_2024}. When larger achievement tests are prepared, the properties of the items (i.e., test questions) need to be understood in advance, and therefore, they are pretested. An item has, under the item response theory (IRT) paradigm \citep{Baker2004}, special characteristics (item parameters) $\boldsymbol{\beta}$ that need to be estimated with as high precision as possible before they are used in a real testing situation. The design problem consists of allocating examinees to $n$ items according to their independent variables called abilities, which are denoted $\theta$. Every examinee can be given at most $d$ items. \cite{Miller_Fackle-Fornius_2024} divided the possible $\theta \in{\rm I\!R}$ into a set of intervals such that an examinee with ability in that interval is given items decided by the optimal design. We can also let the number of intervals be equal to the number of objects/individuals in the general setting.

\cite{Miller_Fackle-Fornius_2024} used the simulated annealing algorithm to find an approximately optimal design. The approach works well, but it is rather time-consuming. 

Another example where incomplete block designs are used is described by \cite{alves2025unifiedbetaregressionmodel}. They model sensory analysis with an incomplete block design where consumers evaluate grape juice formulations for sensory attributes such as color, flavor, aroma, acidity, and sweetness. The study suggested a beta regression model. Even if they did not allocate the treatment based on individual characteristics, the use of a model makes it possible to adjust it to model-based selection of an optimal design.

We compare the algorithms using the example of optimal item calibration in achievement tests by \cite{Miller_Fackle-Fornius_2024}, where D-optimality is the optimization criterion \citep{Atkinsson}. The comparisons are made with respect to the final criterion value, number of iterations, user time, and best-so-far objective in relation to user time. We define four Cases by varying the number of treatments, the size of the subset of treatments for an object, and the number of blocks.

We compare the simulated annealing algorithm, the threshold accepting algorithm, and a deterministic first and best improvement local search algorithm to our directional derivative versions of the first improvement and the chained local search methods using the threshold accepting criterion. 

We conclude that the simulated annealing and threshold algorithm work well for optimizing model-based incomplete block designs. If computing time is not an issue, using simulated annealing with a large number of iterations at every temperature has nice convergence properties. The threshold algorithm produces almost equally good results as simulated annealing. However, our main finding is that our algorithms that utilize directional derivatives reach criterion values close enough to those of the simulated annealing algorithm. This means that we gain a lot of computing time while losing only a small amount of precision in parameter estimates.

The article is organized as follows. Section \ref{Theo} describes the optimal experimental design setup. In Section \ref{algorithms}, we discuss the different optimization algorithms. Section \ref{Eval} describes the setting that we use to compare the different algorithms, and the results are presented in the Results section \ref{Results}. We conclude in Section \ref{Disc} with a discussion. The appendix contains pseudo-code for the algorithms.

\section{Theory}
\label{Theo}

\subsection{Optimization problem}\label{OptProb}
We assume that we can observe a response variable $y$ which depends on independent variables or covariates $\boldsymbol{x}$ with some random measurement error according to a parametric model. In experimental design problems, we can choose the covariates $\boldsymbol{x}$ of the observations to be made from a so-called design region $\chi$. The elements of the design region are called design points. A design $\xi$ is a probability measure on $\chi$ describing which observations we make; see \cite{Atkinsson}, Chapter 9, for a further description. The main objective of the optimal experimental design approach is to find the design $\xi$ that maximizes a function of the Fisher information matrix about the model parameters.

\begin{figure}[h]
\centering
\includegraphics[scale=0.8]{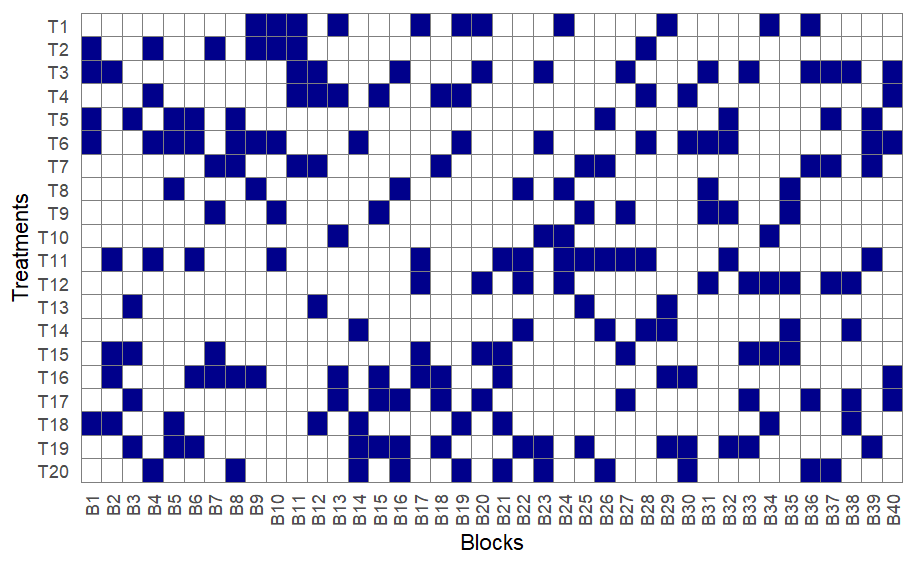}
\caption{An example of a design represented by a binary incidence matrix. The blue cells represent ones and the white cells represent zeros. Note: The R code for the figure was generated using Google Gemini (3.1 Pro, accessed August 2026)}
\label{fig:bin_mat}
\end{figure}

In our case, independent variables are both the treatment $i \in \{1, \dots, n\}$ assigned and individual-specific covariates $\boldsymbol{x}_j \in \tilde\chi = {\rm I\!R}^m$ which are assumed to be known. Formally, the design region is $\chi=\tilde\chi \times \{1,\dots,n\}$, the product space of individual covariates and treatment.
The independent variables can be, e.g., ability of examinees, biomarkers, or disease severity of patients. Our optimal design problem is to allocate individuals to $d$ out of $n$ treatments ($d < n$), using a model-based approach. One way to describe the design is by a binary incidence matrix $(d_{ij})$ of 0s and 1s with treatments in the rows and, say, $J$ individuals in the columns. The cell entry $d_{ij}$ indicates whether that individual receives the specific treatment, i.e.,
$$
d_{ij}= \begin{cases}
    1 \text{, if treatment $i$ is given to subject $j$,}\\
    0 \text{, otherwise.}
\end{cases}
$$
Figure \ref{fig:bin_mat} shows an example of an incidence matrix with $n=20$ treatments and $J=40$ individuals (or blocks), where each individual receives $d=5$ treatments.

Every treatment is associated with a vector of parameters $\boldsymbol{\beta}_i$ that needs to be estimated, and we aim to find a design that allows us to estimate the unknown model parameters with the highest precision.
The problem is now a combinatorial optimization problem where the solution is one of the incidence matrices among all possible ones. The goal is to find an optimal or near-optimal solution according to some criterion, where optimization time is also a factor. The number of different solutions increases rapidly with column and row size, which makes it suitable for using heuristic optimization techniques. 

We assume that the expected value of the response $Y_{ij}$ for treatment $i$ and subject $j$ is a function of $\boldsymbol{\beta}_i$  and $\boldsymbol{x}_j$, $E[Y_{ij}]=f(\boldsymbol{x}_j,\boldsymbol{\beta}_i)$. The general Fisher information matrix for $\boldsymbol{\beta}_i$ for treatment $i$ can be written as
\begin{equation}\label{InfMat}
    M_i = \sum_{j=1}^J F(\boldsymbol{x}_j,\boldsymbol{\beta}_i)\nu_i(\eta) F(\boldsymbol{x}_j,\boldsymbol{\beta}_i)^T d_{ij},
\end{equation} 
where  $\eta = \eta(\boldsymbol{x}_j,\boldsymbol{\beta}_i)$ and $F(\boldsymbol{x}_j,\boldsymbol{\beta}_i)=\frac{\partial \eta(\boldsymbol{x}_j,\boldsymbol{\beta}_i)}{\partial \boldsymbol{\beta}_i}$, see \cite{ATKINSON201481}. For linear and generalized linear models (GLM), $\eta(\boldsymbol{x}_j,\boldsymbol{\beta}_i) = \boldsymbol{\beta_i}^T f(\boldsymbol{x}_j)$ \citep{FedorovLeonov,Atkinsson}. For non-linear or generalized non-linear models (GNLM), $\eta(\boldsymbol{x}_j,\boldsymbol{\beta}_i) = f(\boldsymbol{x}_j,\boldsymbol{\beta}_i)$ \citep{BiedermannWoods, ATKINSON201481, SeberWild}. Further, $\nu_i(\eta)=\nu_i(\eta(\boldsymbol{x}_j,\boldsymbol{\beta}_i)) = \text{Var}\left(\frac{\partial}{\partial \eta} \text{ln} p(Y\mid \eta)\right)$ where $p$ is the density of the data \citep{ATKINSON201481} which equals $\frac{1}{\sigma^2}$ for linear and non-linear models with normally distributed independent errors $\varepsilon_i$ with zero mean and constant variance $\sigma^2$. 

For GLM and GNLM, $\nu(\eta(\boldsymbol{x}_j,\boldsymbol{\beta}_i)) = \left(\frac{\partial \mu}{\partial \eta} \right)^2 / \text{V}(\phi \mu_i)$ \citep{Atkinsson, BiedermannWoods}, where $\mu_i = \text{E}(Y)$, $\text{Var}(Y) = \phi \text{V}(\mu_i)$, and $\phi$ is the dispersion parameter, e.g., $\sigma^2$ for the normal distribution and one for the binomial \citep{Atkinsson}. In the case of known heteroskedastic variance in the linear and non-linear cases, we can write $\nu(\eta(\boldsymbol{x}_j,\boldsymbol{\beta}_i))= \frac{1}{\sigma^2(\boldsymbol{x}_j)}$.

The objective is to estimate all the parameter vectors $\boldsymbol{\beta}_i$ with good precision, which means that the parameter vector of interest is $\boldsymbol{\beta}=(\boldsymbol{\beta}_1^T,\dots,\boldsymbol{\beta}_n^T)^T$. If we can assume independence between treatments, the total information matrix for $\boldsymbol{\beta}$ is block diagonal $M = \text{diag}(M_1,\dots,M_n)$.

If the number of individuals $J$ is too big, it would be too demanding for the optimization to find the best treatment combination for each individual. Furthermore, it is reasonable that individuals with similar characteristics $\boldsymbol{x}$ receive the same treatment combination. We introduce, therefore, blocks of individuals with similar characteristics that are supposed to be allocated to the same treatment combination. If $\boldsymbol{x}=x$ is one-dimensional, the group of individuals with independent variable $x$ below some cutoff would be grouped into the first block, individuals with $x$ between this cutoff and some higher cutoff would be in the second group, and so on; see \cite{Miller_Fackle-Fornius_2024}. Generally, including multidimensional $\boldsymbol{x}$, we divide the space of independent variables into $b$ disjoint subsets or blocks $S_j, j=1, \dots, b$. The design can now be described by a $n\times b$ incidence matrix $(d_{ij})_{i=1,\dots,n; j=1,\dots,b}$.

If $J$ is large, we can describe the population of available individuals by a probability density $g(\boldsymbol{x})$ over the independent variables. 
In this case, the information matrix can be written as
\begin{eqnarray}
M_i & = & \sum_{j=1}^b \int F(\boldsymbol{x},\boldsymbol{\beta}_i)\nu_i(\eta) F(\boldsymbol{x},\boldsymbol{\beta}_i)^T d_{ij} g(\boldsymbol{x})d\boldsymbol{x} \nonumber \\
 & = & \int F(\boldsymbol{x},\boldsymbol{\beta}_i)\nu_i(\eta) F(\boldsymbol{x},\boldsymbol{\beta}_i)^T h_i(\boldsymbol{x})d\boldsymbol{x}, \label{InfMat_dens}
\end{eqnarray} 
where $h_i(\boldsymbol{x})$ describes the sub-density of individual assigned to treatment $i$ defined as 
\begin{equation}\label{subden}
h_i(\boldsymbol{x})= \sum_{j=1}^{b}d_{ij}\boldsymbol{1}\{\boldsymbol{x}\in S_j\}g(\boldsymbol{x}).
\end{equation} 

The aim is now to find an optimal design that maximizes the information in some appropriate way by optimizing a functional $\Psi(M)$ of the information. We choose here D-optimality as the optimization criterion \citep{Atkinsson}, which means to maximize the determinant of $M$ or, equivalently, to minimize 
\begin{equation}\label{Dopt_log}
\Psi(M)=-\text{log\{det}(M)\} = -\sum_{i=1}^n\text{log\{det}(M_i)\}.\end{equation}

When the function $f(\boldsymbol{x}_j,\boldsymbol{\beta}_i)$ is non-linear, the information matrix depends on the parameters $\boldsymbol{\beta}_i$ to be estimated. Therefore, we must assign initial values to the parameters $\boldsymbol{\beta}_i$ in case of non-linearity. When we then optimize the information matrix for these initial parameter values, this is referred to as a locally optimal design \citep{Atkinsson}.

\subsection{Directional derivative}\label{ddriv}

Since the directional derivative is used in one of our proposed algorithms, we here provide a definition. The directional derivative instructs on how the information of one design $\xi$ changes in the direction of another design $\lambda$. The directional derivative is defined as
\begin{equation}
    F_{\Psi}(\xi,\lambda) = \lim_{\alpha \rightarrow 0}\frac{1}{\alpha}\left( \Psi(M((1-\alpha)\xi + \alpha\lambda)) - \Psi(M(\xi))\right),
\end{equation}
see, e.g., \cite{Silvey,Hassan2019}.
If we let $\lambda$ be the measure $\delta_{(\boldsymbol{x}, i)}$, with unit weight at design point $\boldsymbol{x}$ for treatment $i$, we can quantify the criterion change when adding a small number of observations at $\boldsymbol{x}$ for treatment $i$.

\section{Optimization algorithms}\label{algorithms}

The design space is the set of all feasible experimental designs that can be represented by an $n\times b$ incidence matrix $\mathbf{D}$ such that every column $j$ must contain $\leq d$ ones and consequently $\geq n-d$ zeros. 
Formally, the design space is
\begin{equation}\label{desSpace} \mathcal{D}=\biggl\{\mathbf{D} \in \{0,1\}^{n \times b} : \sum_{i=1}^{n}d_{ij} \leq d~ \forall j=1,\dots, b \biggr\}.\end{equation} 
Since D-optimality is a monotonic optimality criterion, adding observations does not make the designs worse. Therefore, we need only to consider the subset $\mathcal{D}'$ where each individual receives exactly $d$ treatments, $\sum_{i=1}^{n}d_{ij} = d$. The total number of potential designs is then given by
\begin{equation}\label{totDes} \binom{n}{d}^b = \left(\frac{n!}{d!(n-d)!}\right)^b.\end{equation}

The objective is now to find the solution $\mathbf{D}$ over $\mathcal{D}'$ that minimizes the function $\Psi(M) = -\sum_{i=1}^n\text{log[det}( M_i)]$ which is a combinatorial optimization problem. 

Approaches for solving combinatorial optimization problems are either exact or approximation methods. When the problem becomes too complex, it is common to use heuristic methods \citep{Reeves, Talbi, Taillard2023} with approximate solutions. Common heuristic methods in combinatorial optimization are, e.g., simulated annealing, tabu search, variable neighborhood search, and iterated local search \citep{Bianchi2009Survey, Baghel2012}, which can be classified as local search methods. Lesser-known local search methods are the threshold accepting algorithm \citep{DUECK1990161} and chained local optimization. The threshold accepting algorithm has been shown to outperform simulated annealing in both solution and computing time. The chained local optimization restricts the sampling in the simulated annealing algorithm to locally optimal configurations only, using a local search algorithm. The method is, therefore, sometimes called a large-step Markov chain or basin-hopping algorithm.

Exact methods provide, in theory, a global optimal solution. Examples of exact methods are complete enumeration \citep{Papadimitriou1998}, dynamic programming \citep{Bellman1962}, integer programming \citep{Nemhauser1999}, and branch-and-bound methods \citep{LandDoig1960}. In the examples we will explore, the problems are (often) too large and complex for exact methods. 

Because of the large number of solutions, we will propose using different versions of local search methods. The number of possible solutions will get very big with just a few treatments, which would make an exhaustive search and other exact methods infeasible to use. 

\subsection{Local search algorithms}\label{LocalSearch}
Local search is a class of heuristic models for finding an optimum in computationally hard problems. The algorithms begin at a starting point and move from one solution to another by local changes. The search continues until an optimal or near-optimal solution is found. A key concept in local search is neighborhoods that are close to the current solution according to some definition. See, e.g., \cite{AartsLenstra2003}. Closeness can mean that they are easily computed. In complex problems, the iterative search is, however, likely to end up in a local optimum, often too far from the global optimum \citep{Kong2021,Lourenco2003}.

A neighborhood $U(s,\phi)$ of a solution $s$ is a set of solutions that is reachable by an operation $\phi$. An operation might be removing or adding an object from the solution, or it might be exchanging an object within the solution with an object outside. If a solution $s$ is better than all other solutions in its neighborhood $U(s,\phi)$, then it is a local optimum regarding this neighborhood. See \cite{Reeves,AartsLenstra2003} for a further description of neighborhood structure in local search. 

Two types of local search algorithms are the first and best improvement algorithms. The first improvement algorithm searches within the neighborhood of the current solution until the first improvement is found, whereas the best improvement algorithm updates the solution with the best improvement within the neighborhood. In experimental optimal design, these are called the Fedorov exchange algorithm \citep{Fedorov} and the \cite{CookNachtsheim1980} version of the exchange algorithm. 

The exchange algorithms of \cite{Fedorov} and \cite{CookNachtsheim1980} are used in classical experimental optimal design problems, where the logarithm of the determinant (D-optimality) is concave on the cone of positive‑definite matrices \citep{Silvey}. This becomes a convex maximization problem where a global optimum can be found. We are, however, optimizing an incidence matrix with column constraints. The objective function is still the logarithm of the determinant of a block-diagonal information matrix, but without the global concavity of the objective function. This produces many local optima, which means that our algorithms must be able to handle this.

Two heuristic techniques for discrete search spaces that avoid local optima too far from the global optima are simulated annealing \citep{KirkpatrickGelattVecchi, Reeves} and threshold accepting algorithms.

Simulated annealing starts with a random solution and then chooses a new candidate from a neighborhood of the current solution. The new candidate solution is accepted with a probability that depends on the objective function values of the solutions and a control parameter called temperature. Assume $\Delta E$ is the difference in objective value between a new and old solution. When minimizing, an improvement $(\Delta E <0)$ of the objective function is always accepted; otherwise, the new candidate solution is accepted with the acceptance probability $\text{exp}(\frac{-\Delta E} {t} ) $, called the Metropolis criterion. The temperature $t>0$ will gradually decrease, making it less probable to accept a worse solution over time. The algorithm explores a wider range of solutions (including worse solutions) earlier in the process.

Besides the decision of the initial temperature and definition of the neighborhood, the number of iterations before the next temperature reduction needs to be determined. A reduction function is also needed that determines the magnitude of the temperature decrease in every cycle.

A strength of simulated annealing is that it has been proved \citep{CruzDorea1998, Mitra-etal1986, Kirkpatrick} that it converges asymptotically to the optimal solution. Unfortunately, it requires exponential time. But if the algorithm is allowed to explore a large search space before every temperature reduction, the solution will be approximately good \citep{Reeves}. 

The threshold accepting algorithm can be seen as a simpler version of the simulated annealing algorithm. The difference between the algorithms lies in the acceptance criterion. The threshold accepting algorithm accepts every new solution that is slightly worse than the preceding solution. If we assume $\Delta E$ is the difference in objective value between a new and old solution, a solution is always accepted if $\Delta E < t$ in the case of minimization, and $t$ is the acceptance threshold. Simulated annealing accepts worse solutions (large or small) with a probability. 

An advantage of the algorithm is that there is no need for computing probabilities or making random decisions. At least for some problems, it shows better results in terms of final objective function value and less computing time \citep{DUECK1990161}. We will, therefore, consider the algorithm as a comparison to simulated annealing.

A method called chained local optimization \cite{Martin1996} combines simulated annealing with local search methods for combinatorial optimization problems. The idea is to combine local search techniques with simulated annealing so that all solutions that are considered by the Metropolis criterion are local optima. The idea is to take advantage of local search methods to overcome large barriers in the solution space. The algorithm works as follows. Apply a (possibly deterministic) local search algorithm to a start solution. Then apply a perturbation to the solution and run the local search algorithm again. If the new local optimum is worse, apply the Metropolis criterion from simulated annealing that decides whether this new local optimum is rejected or not; otherwise, just accept the new solution. If it is rejected, return to the start. This algorithm is described very well in Figure 1 of \cite{Martin1996}.

We will use the same idea, but instead of using the Metropolis version of the accept/reject criterion, we will use the threshold accepting criterion. As a local search algorithm, we use the first improvement algorithm using directional derivatives described in Section \ref{neighExAlg} as the algorithm that moves from one local optimum to another.

\subsection{Description of the algorithms used}\label{neighExAlg}
We will describe six algorithms that we compare for optimizing the combinatorial problem we have described. The algorithms are the simulated annealing (SA) algorithm, the threshold accepting (TA) algorithm, the best improvement (BI) exchange algorithm, the first improvement (FI) algorithm, the first improvement algorithm using directional derivatives (FIdd), and the chained local optimization algorithm with the FIdd algorithm and the threshold accepting criterion (TAdd). 

A general principle for an iteration in all algorithms is: Choose a column (block) $v \in \mathcal{B} = \{1,\dots,b\}$ according to some rule $\delta$ (which can be a random or systematic choice); do then an exchange of row-entries in the incidence matrix $\mathbf{D} \in \mathcal{D}^{'}$. We define the row operation as the function $\phi_k$ that exchanges $k$ of the ones with $k$ of the zeros in the chosen column vector $\mathbf{d}_v$, where $k \in \{1,\dots,\text{min}(d,n-d)\}$. A $k$-exchange operation $\phi_k: \{0,1\}^n \to \{0,1\}^n$ is $\phi_k(\mathbf{d}_v)=\mathbf{d}^{'}_{v}=(d^{'}_{i1},\dots,d^{'}_{in})^T$ with
\begin{equation}
   \sum_{i=1}^{n} d^{'}_{iv} = d \text{ and } ||\mathbf{d}_v -  \mathbf{d}^{'}_v|| = 2\cdot k,
\end{equation} 
where $||\mathbf{d} -  \mathbf{d'} || 
$ is the $L^1$ distance. It is also equal to the Hamming distance \citep{Hamming1950}, which is the number of positions where the values of the positions differ. After a column has been selected, the number of possible $k$-exchanges is $\pi_k = \binom{d}{k} \binom{n-d}{k}$.

All algorithms start with an incidence matrix $\mathbf{D}_{start}$. For the SA, TA, and TAdd algorithms, $\delta$ is the function choosing a column in $\mathbf{D}_{start}$ randomly with equal probability. The other three algorithms choose the first column. In all algorithms, we will only consider operations exchanging a 1 and a 0, that is, an operation $\phi_1$.

It has turned out that when using only exchanges restricted to a single column, the algorithms often remain in local optima. The reason is that too few observations in one row might lead to a bad D-optimality criterion. Then removing a specific 1-entry in a chosen column might never be done even if exchanging the 1 with another 0 in that row would improve the criterion. We therefore also use another exchange operation, $\tilde \phi_1:\mathcal{D} \to \mathcal{D}$, involving two columns. We choose column $v$ as before and do an exchange $\phi_1$ in this column; we balance this exchange out at the row level by looking for another column $v'$ which has a 0 and a 1 in opposite rows compared to column $v$:
\begin{equation}
\label{eq_phitilde}
   v' \in \mathcal{B}\backslash \{ v\},
   \sum_{i=1}^{n} d^{'}_{iv} = \sum_{i=1}^{n} d^{'}_{iv'} = d, ||\mathbf{d}_v -  \mathbf{d}^{'}_v|| = ||\mathbf{d}_{v'} -  \mathbf{d}^{'}_{v'}|| = 2, \mathbf{d}_v -  \mathbf{d}^{'}_v + \mathbf{d}_{v'} -  \mathbf{d}^{'}_{v'} = \mathbf{0}.
\end{equation} 
In each iteration in all algorithms, we therefore perform both an operation $\phi_1$ and an operation $\tilde\phi_1$ (with the same $v$). The criterion value $\Psi(M)$ is calculated for both operations, and the operation with the higher criterion value is used for further evaluation in the algorithms. If a $v'$ cannot be found according to (\ref{eq_phitilde}) (in practice, this happens only rarely), only the $\phi_1$ operation is used.

In the \textbf{SA algorithm}, the current incidence matrix is updated with the proposed matrix based on $\phi_1$ or $\tilde \phi_1$ if $\Psi(M)$ is improved; otherwise, it is updated with some probability depending on the temperature of SA. The algorithm is run for a predetermined number of runs for every temperature and a predetermined number of times before every lowering of the temperature according to a cooling scheme. 

The \textbf{TA algorithm} works in the same way as SA, but 
the new solutions are updated if the difference between the new and the old solution is smaller than a predetermined threshold. The algorithm is run for a predetermined number of runs, and the threshold value can be lowered a predetermined number of times, but this is not necessary.

The \textbf{BI algorithm} calculates the criterion value $\Psi(M)$ for all possible exchanges between 1's and 0's based on $\phi_1$ and updates the exchange that improves $\Psi(M)$ the most. For the BI algorithm, we consider an exchange based on $\tilde \phi_1$ according to (\ref{eq_phitilde}) only for the $\phi_1$-exchange that improves the criterion most, but not for all possible exchanges in the column.

The \textbf{FI algorithm} orders all combinations of 1s and 0s and iterates through the combinations one at a time. It stops iterating as soon as $\Psi(M)$ is improved.

The \textbf{FIdd algorithm} orders all combinations of 1s and 0s in decreasing order based on the differences of the directional derivative values that correspond to the 1s and 0s. 
We now have a matrix with $n \times b$ directional derivative values along with the incidence matrix $\mathbf{D}$. The algorithm iterates through the combinations 
one at a time. It stops iterating as soon as $\Psi(M)$ is improved. 

The updated incidence matrices are now $\mathbf{D}_{current}$, and the operation $\delta$ is applied again with a random selection of columns for SA and by selecting the second column for the others. 

The algorithms BI, FI, and FIdd stop if no improvements have been found for $b$ columns. The algorithms are formally described in Algorithm \ref{Alg:BI} and Algorithm \ref{Alg:FIFIdd} in the appendix. 

The final algorithm we suggested is the \textbf{TAdd algorithm}. It will run the FIdd algorithm on a starting solution and perturb the intermediate solution. FIdd is run again until a new local optimum is found. If that solution is better than the starting solution, it will be accepted. And if it is worse, but not too bad according to the threshold, it will be accepted. If it is rejected, it returns to the starting solution. The threshold can be lowered by some schedule or can be held constant.

Without the threshold criterion, the algorithm would just be an iterated local search algorithm, which means that, after finding a local optimum, the solution is perturbed and the local search algorithm is performed with the perturbed solution as the starting value \citep{Lourenco2003}. Perturbation in this context means that, after finding a local optimum, we randomly select $l$ columns of the matrix $\mathbf{D}$ and $m$ 0s and 1s in each of those columns. The selected 0s and 1s are exchanged in the $l$ columns, and the exchange algorithm starts again with the perturbed version of $\mathbf{D}_{pert}$ as the starting binary matrix. The general idea behind perturbation is to escape local optima and find solutions with an objective function value closer to the optimum. See, e.g., \cite{Talbi} and \cite{Lourenco2003} for descriptions of perturbation methods in optimization. 

In the algorithms described, we only consider 1-exchange neighborhoods. If we consider, e.g.,  $n = 60$ and $d=10$, the number of possible $1$-exchanges is $\binom{10}{1}\binom{50}{10} = 500$. The number of possible $2$-exchanges is $\binom{10}{2}\binom{50}{2} = 45 \cdot 1225 = 55125$, which means that a lot of solutions are not reachable with only $1$-exchanges. However, 

it can be seen that the set of all $n-\text{bit}$ vectors with exactly $d$ ones is the vertex-set of the Johnson graph $J(n,d)$ whose edges are precisely those $0-1$ exchanges. The fact that $J(n,d)$ is connected means that all solutions can be reached from any other by a path of 1-exchanges \citep{Diestel2021}. It means that it is theoretically possible to reach all solutions from a given matrix $\mathbf{D}$. By using only 1-exchanges, we simplify the algorithms a lot.

\subsection{Theoretical justification of the use of directional derivatives}

The idea of using directional derivatives in the first improvement algorithm is to make smaller and smoother steps in the criterion values compared to the first and best improvement algorithms. The directional derivative version makes the first improvement algorithm do wiser exchanges, which gives it better properties. \cite{Hassan2019,ULHASSAN2021107177} used directional derivatives in the context of item calibration, but they had a continuous design space.

The \text{BI algorithm} not only needs to calculate the criterion value for every possible exchange, but also takes steps that are too large. It makes it less sensitive to the surface of the search space, especially after a while,  and it will likely end up in a local optimum too far away from the global optimum. 

In many of the designs we use, the criterion function is locally flat, which means that many small exchanges don’t change the criterion much. The \text{FI algorithm} will spend time testing and rejecting many of these small, neutral exchanges. The \text{FIdd algorithm}, however, prioritizes exchanges with the largest difference in directional derivatives, effectively avoiding flat regions.

The \text{FIdd algorithm} is a discrete version of a continuous gradient-based algorithm. The algorithm, therefore, inherits theoretical justification from continuous theory. It also mimics the logic of the Fedorov exchange algorithm \citep{Fedorov} for approximate designs, but adapted to a discrete setting.

\section{Evaluation of the method}\label{Eval}
We will use an example from item calibration used in educational assessments to evaluate the algorithms. \cite{Miller_Fackle-Fornius_2024} describe a situation where $n$ test items need calibration before being used in a real testing situation. Calibration here means estimating item parameters $\boldsymbol{\beta}$ with as high precision as possible. The problem is to allocate the examinees according to their ability $\theta$ in an optimal way. The relation between the item parameters $\boldsymbol{\beta}$ and the examinee's ability $\theta$ depends on the chosen model. We use the two parameter logistic (2PL) model \citep{Baker2004}, which can be written as follows

\begin{equation}\label{2PL}
    p_i(\theta) = \frac{1}{1 + \text{exp}(-\beta_{i1}(\theta -\beta_{i2}))},
 \end{equation} 
 where $\beta_{i1}$ is called discrimination- and $\beta_{i2}$ difficulty-parameter of the item in this context.
 The 2PL logistic model is a GLM. Therefore, the information matrix for $\boldsymbol{\beta}_i = (\beta_{i1},\beta_{i2})$ is
\begin{equation}\label{InfMat2}
M_i = \int p_i(\theta)(1-p_i(\theta)) \left(\frac{\partial \eta(\theta,\boldsymbol{\beta}_i)}{\partial \boldsymbol{\beta}_i} \right)\left(\frac{\partial \eta(\theta, \boldsymbol{\beta}_i)}{\partial \boldsymbol{\beta}_i} \right)^T h_i(\theta)d\theta,
\end{equation}
where $\nu_i(\eta) = p_i(\theta)/(1-p_i(\theta))$ and $\eta(\theta, \boldsymbol{\beta}_i) = \text{log}\{p_i(\theta)/(1-p_i(\theta))\}$, see \cite{Hassan2019} and \cite{ULHASSAN2021107177} for a further description. 

As described by \cite{Miller_Fackle-Fornius_2024}, we assume that an examinee can only manage answering at most $d$ items and that we have $b$ test versions. We divide the population of examinees into $b$ subpopulations of approximately equal size. Then we assign each subpopulation to one test version. The first test will be assigned to examinees with abilities smaller than $\theta_1$, the second to all examinees with a value lower than $\theta_2$ but higher than $\theta_1$, and so on. 

We let the discrimination parameters $\beta_{i1}=1$ for all items and the difficulty parameters $\beta_{i2}$ is equidistant between $-2$ and $2$. We use the D-optimality criterion  $\Psi(M)=-\sum_{i=1}^n\text{log}\{\text{det}( M_i)\}$ and the directional derivative is as described in \cite{Hassan2019}:
\begin{equation}
    F_{\Psi}(\theta,i) = 2 - p_i(\theta)(1-p_i(\theta))\left[(\theta - \beta_{i2})~ (-\beta_{i1}) \right] M_i^{-1}\left[(\theta - \beta_{i2})~ (-\beta_{i1}) \right]^T.
\end{equation} See \cite{Miller_Fackle-Fornius_2024} for a detailed description of the setup.

To be able to notice how the methods change with problem size, we compare the algorithms using four cases, which are combinations of item numbers $n$ and number of treatments per block $d$, while keeping $b=40$ constant; see Table \ref{tab:4cases}. 
\begin{table}[htbp]
    \centering
    \begin{tabular}{|c|cc|}
               & $n$ & $d$ \\ \hline 
      Case I   & 5   & 2   \\
      Case II  & 10  & 4   \\
      Case III & 30  & 8   \\
      Case IV  & 60 & 10   \\ \hline
    \end{tabular}
    \caption{Four cases to compare the algorithms}
    \label{tab:4cases}
\end{table}

For the SA algorithm, a good choice for the starting temperature is a value that produces an acceptance ratio smaller but close to one \citep{KirkpatrickGelattVecchi}. For simplicity, we chose the starting temperature to be $t=1$ for all cases. We choose the cooling factor to be $\alpha = 0.85$, such that the temperature decreases as $t_{k +1} = \alpha \cdot t_k$, and will be close to zero in the end. The number of iterations per temperature level is set to 50000 for Case I and 150000 for Case II-IV, and the algorithm iterates through 50 different temperatures. The iteration numbers are chosen such that we believe the sequence of trials is close to equilibrium before every temperature reduction and that the final temperature is close to zero.

For the TA algorithm, we choose the threshold between $t= 0.005 - 0.01$ for the different cases, and we reduced the threshold by multiplying by the factor $\gamma = 0.95$. The numbers were chosen by trial and error and could probably be set in a more efficient way. The number of iterations before every threshold reduction is the same as for SA. Both the TA and SA algorithms are stopped if more than ten outer iteration runs do not improve the best-so-far value of the objective function.

As described in Section \ref{neighExAlg}, we perturb the solution after every local search in the TAdd algorithm. The perturbation is made by exchanging $l$ zeros and ones in $m$ columns. We have, by testing different combinations of hyperparameters $l$ and $m$, seen that larger perturbations are needed for larger values of $n$ and $d$. So far, we used $l=2$ and $m=2$ all combinations of $n$ and $d$ except for $n=60$ and $d=10$ where we used $l=5$ and $m=2$. Since every local search in TAdd is one round of the FIdd algorithm, the total number of runs is set to $n_{t} =50$. Since it is desirable for the threshold $t$ to be close to zero at the end of the algorithm, we choose to decrease $t$ as $t\cdot\left(1- (\frac{i}{n_t})\right)$, where $i \in (1,\dots, n_t)$. The starting threshold accepting value for the TAdd algorithm is set to $0.002 -0.02$, and the algorithm is terminated if no improvements have been made for more than ten local search rounds.

We run every algorithm 10 times with a random binary starting matrix $\mathbf{D}_{start}$, the same between models but different between iterations. We present the results as tables in a similar way as in \cite{GARCIARODENAS2020106844} and evaluate the mean of final criterion values, mean number of iterations, and mean CPU time in seconds. To see the variation, we also display box plots of the final criterion values for all algorithms and cases.

We also present graphs of the best-so-far values compared to user CPU time in seconds. Since the lengths of the best-so-far outputs differ between runs and between methods, we choose a common time grid and interpolate the objective value with a step function. The objective value at any point on the new time grid is $\tilde{\Psi}_t = \text{min}(\Psi_{t_i}: t_i \leq t)$, where $i$ is the iteration. Then we take the mean at every new time step across all methods and plot them with the CPU time on a logarithmic scale.

\section{Results}\label{Results}
Table \ref{tab:all_cases} shows that for Case I, SA, TA, and TAdd have the same mean criterion, but SA has a substantially longer running time than TA and especially TAdd. The BI, FI, and FIdd algorithms produce solutions very close to the former ones in terms of $\mathrm{RE}_{\mathrm{D}}$. But the latter are a lot faster. For Case II, SA is slower but produces a slightly lower criterion value compared to TA and TAdd. The BI, FI, and FIdd algorithms are still a lot faster.

For Case III, the TA algorithm produces a lower mean criterion value compared to SA, but at almost the same CPU time. The criterion values of BI, FI, and FIdd are still relatively close to the values of SA and TA. However, the mean criterion value of BI is now higher in comparison. A similar pattern can be seen for Case IV.

It can also be seen that the CPU time is similar for FIdd compared to FI and for Cases I and II. For Cases III and IV, the CPU time for FIdd is lower. This indicates that the dd-version of the algorithm does not benefit in the smaller cases due to the extra time of computing the directional derivative matrix.

\begin{table}[phtb]
\centering
\caption{Mean of the function $\Psi$, CPU user time in seconds CPU (s), and number of iterations.}
\label{tab:all_cases}

\subfloat[Case I: $n=5$, $b=40$, $d=2$]{
   \begin{tabular}{l|cccc}
\hline
 Algorithm & $\Psi$ & CPU (s) & Iter.  \\ \hline
SA & 20.61369 &  175.47 & 2500000 \\
TA & 20.61369 &  43.95 & 2500000\\
BI & 20.61481 &  0.10 & 200\\
FI & 20.61754 &  0.26 & 1175.5 \\
FIdd &  20.61439 &  0.28 & 792.2 \\
TAdd & 20.61369 &  3.50 & 10476.7 \\
\hline
\end{tabular}
}
\qquad

\subfloat[Case II: $n=10$, $b=40$, $d=4$]{
  \begin{tabular}{l|cccc}
  \hline
  Algorithm & $\Psi$  & CPU (s) & Iter.  \\ \hline
  SA &  40.54993 & 616.5& 7500000 \\
  TA & 40.55019 & 356.70 & 7500000 \\
  BI & 40.56043& 0.27& 268 \\
  FI &  40.55622& 1.65 &  6604.1 \\
  FIdd & 40.55750& 1.43 & 4229.2\\
  TAdd &40.55052&  22.93& 71165.2 \\
  \hline
  \end{tabular}
} 
\qquad
\subfloat[Case III: $n=30$, $b=40$, $d=8$]{
  \begin{tabular}{l|cccc}
  \hline
  Algorithm & $\Psi$ & CPU (s) & Iter. \\ \hline
  SA & 143.4321 & 1288.06 & 7500000 \\
  TA & 143.4317 & 1254.19 & 7500000  \\
  BI & 143.6341 &7.00 & 556.0\\
  FI & 143.4663 & 38.28 & 99453.3  \\
  FIdd & 143.4579 &19.40 & 37815.4\\
  TAdd & 143.4405& 444.40& 915661.8\\
  \hline
  \end{tabular}
} 
\qquad
\subfloat[Case IV: $n=60$, $b=40$, $d=10$]{
  \begin{tabular}{l|cccc}
  \hline
  Algorithm & $\Psi$ & CPU (s) & Iter.  \\ \hline
  SA & 341.9432 &  1703.90 & 7500000  \\
  TA & 341.9481  & 1642.42  & 7500000 \\
  BI & 343.6581&  36.11 &732 \\
  FI &341.9904 &   140.98  &387118.3 \\
  FIdd & 342.0454&  100.88& 218004.7\\
  TAdd & 341.9652 & 1149.64 &2618184.9 \\
  \hline
  \end{tabular}
}
\end{table}

\begin{figure}[h]
\centering
\includegraphics[scale=0.8]{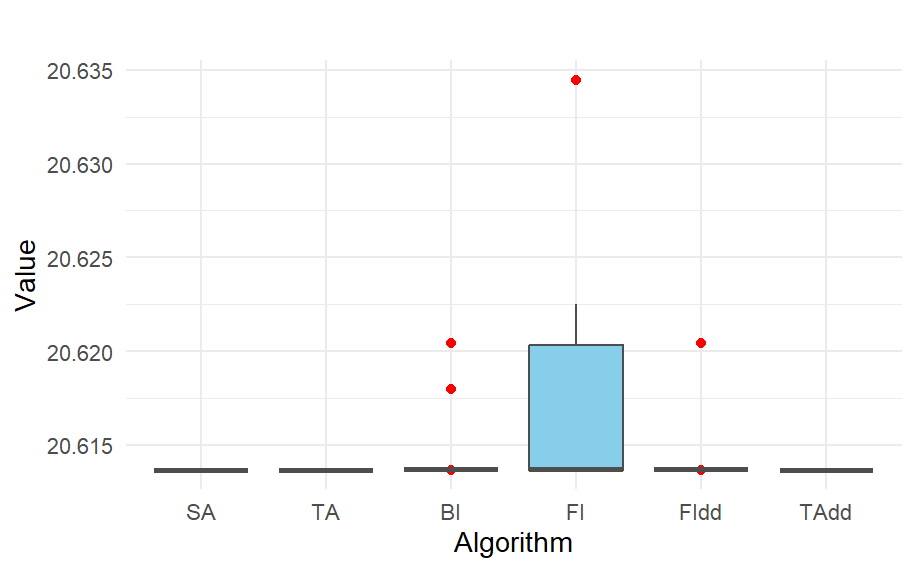}
\caption{Box plots over the 10 final criterion values for all methods for Case I, $n=5$, $b=40$, and $d=2$}
\label{fig:boxplot_5_40_2}
\end{figure} 

\begin{figure}[h]
\centering
\includegraphics[scale=0.8]{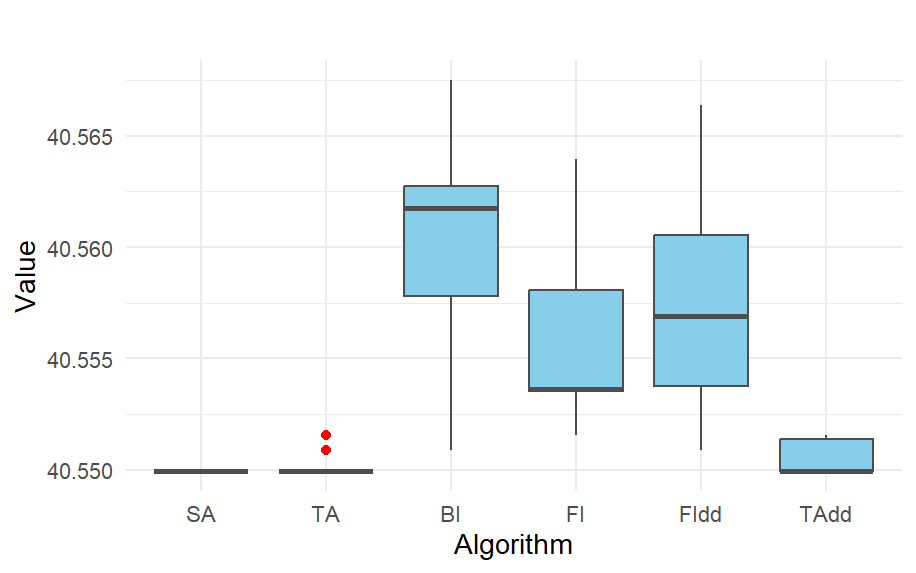}
\caption{Box plots over the 10 final criterion values for all methods for Case II, $n=10$, $b=40$, and $d=4$}
\label{fig:boxplot_10_40_4}
\end{figure}

\begin{figure}[h]
    \centering

    \begin{subfigure}{\textwidth}
        \centering
        \includegraphics[scale=0.85]{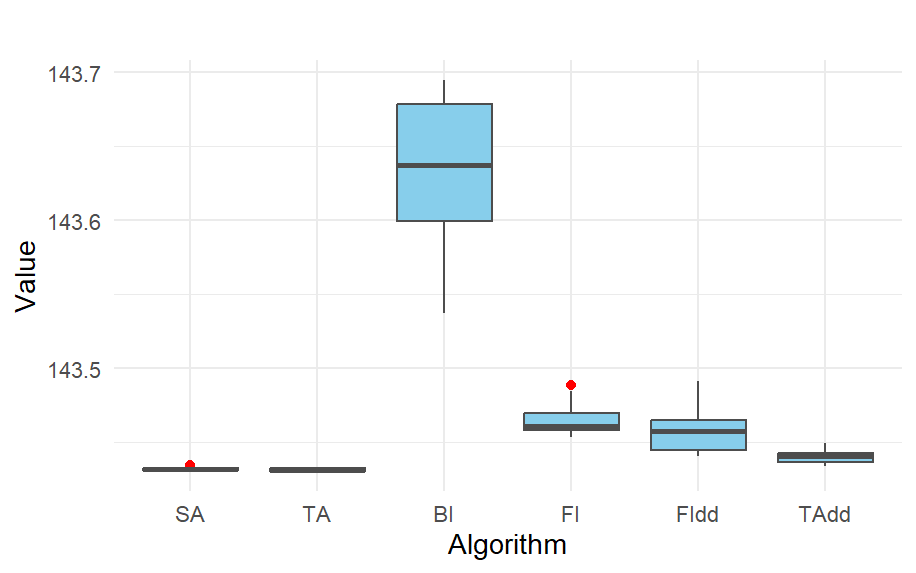}
        \caption{All methods}
        \label{fig:boxplot_30_40_8_BI}
    \end{subfigure}

    \vspace{1em} 

    \begin{subfigure}{\textwidth}
        \centering
        \includegraphics[scale=0.85]{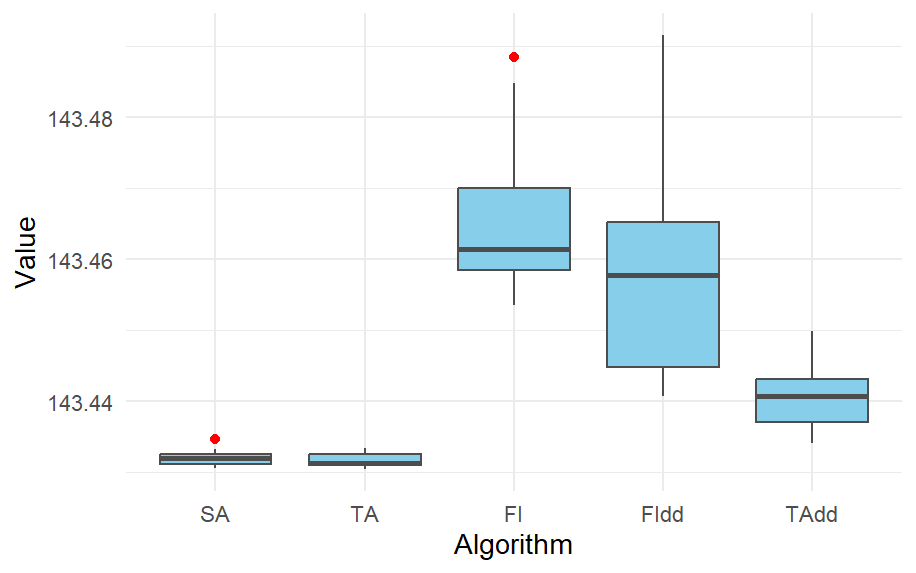}
        \caption{Without BI}
        \label{fig:boxplot_30_40_8_NoBI}
    \end{subfigure}

    \caption{Box plots over the 10 final criterion values for Case III, $n=30$, $b=40$, and $d=8$.}
    \label{fig:Boxplot_30_40_8}
\end{figure}

\begin{figure}[h]
    \centering

    \begin{subfigure}{\textwidth}
        \centering
        \includegraphics[scale=0.85]{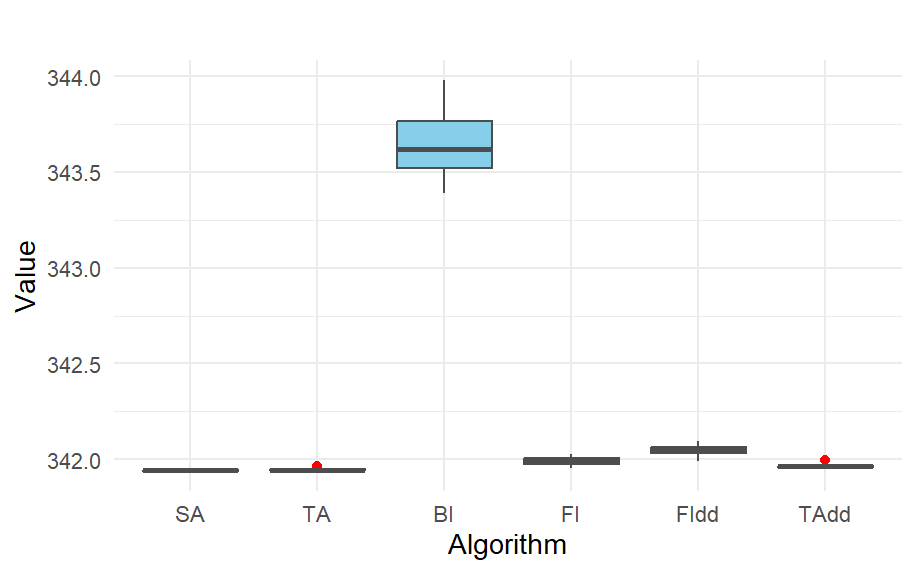}
        \caption{All methods}
        \label{fig:boxplot_60_40_10_BI}
    \end{subfigure}

    \vspace{1em} 

    \begin{subfigure}{\textwidth}
        \centering
        \includegraphics[scale=0.85]{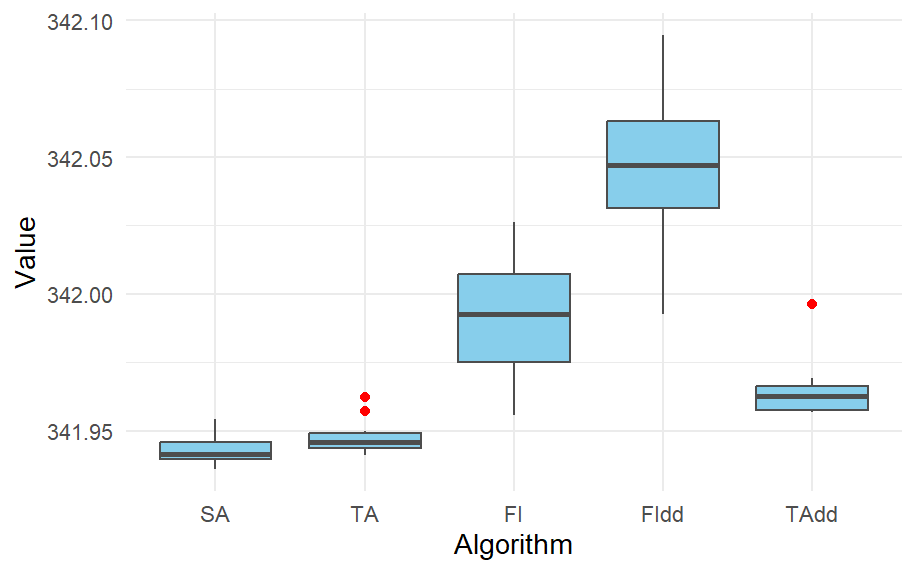}
        \caption{Without BI}
        \label{fig:boxplot_60_40_10_NoBI}
    \end{subfigure}

    \caption{Box plots over the 10 final criterion values for Case IV, $n=60$, $b=40$, and $d=10$.}
    \label{fig:Boxplot_60_40_10}
\end{figure}

Figures \ref{fig:boxplot_5_40_2} to \ref{fig:boxplot_60_40_10_NoBI} show that the FI algorithm has the most variation in the final criterion values for Case I. For Case II, the BI, FI, and FIdd algorithms have the largest variation. The BI algorithm has the largest variation for Case III, followed by FIdd, FI, and TAdd. This pattern is repeated for Case IV, with the difference that the variation for TAdd is just somewhat larger than for SA and TA. The SA and TA algorithms have low variation for all cases. 

\begin{figure}[h]
    \centering

        \includegraphics[scale=0.8]{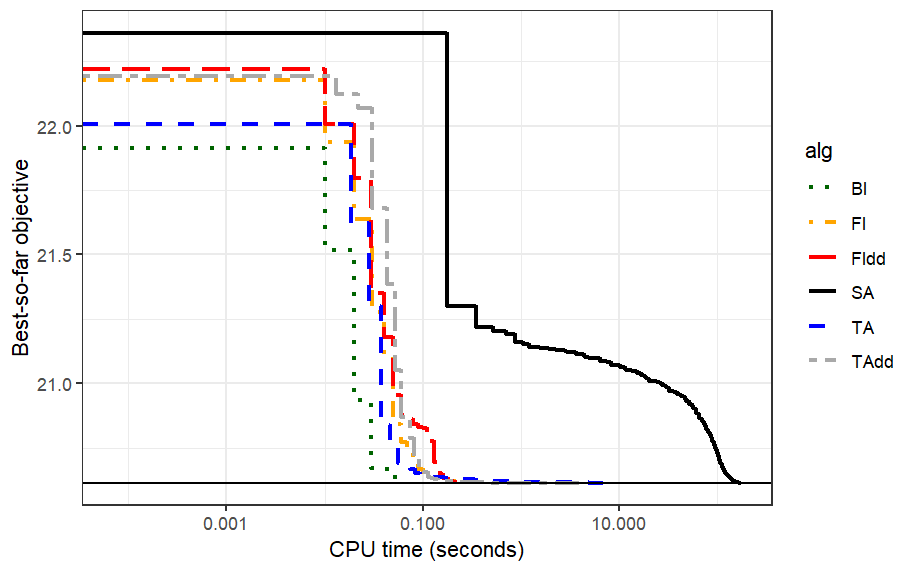}
        \caption{Best-so-far objective by logarithm time for Case I}
        \label{fig:bsf_5_40_2}

\end{figure}

\begin{figure}[h]
    \centering

        \includegraphics[scale=0.8]{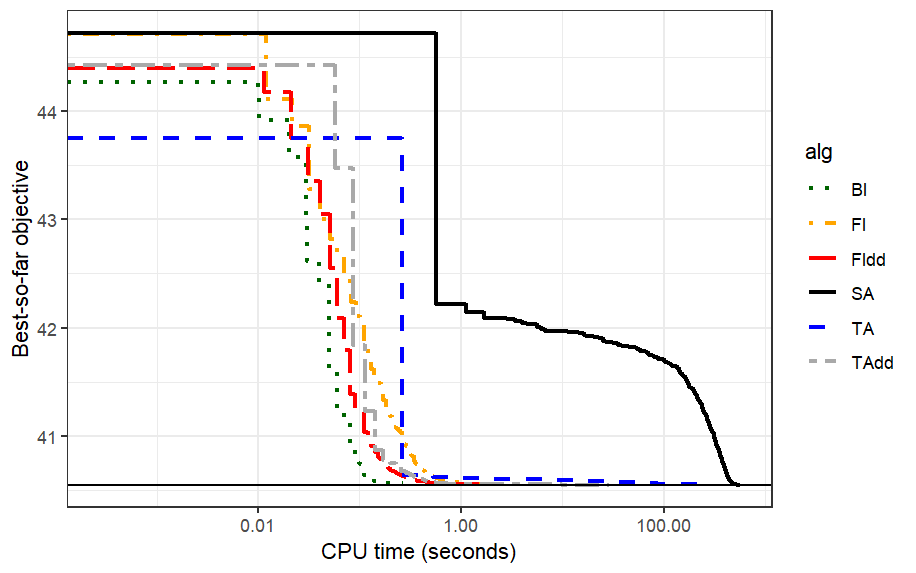}
        \caption{Best-so-far objective by logarithm time for Case II}
        \label{fig:bsf_10_40_4}

\end{figure}

\begin{figure}[h]
    \centering

        \includegraphics[scale=0.8]{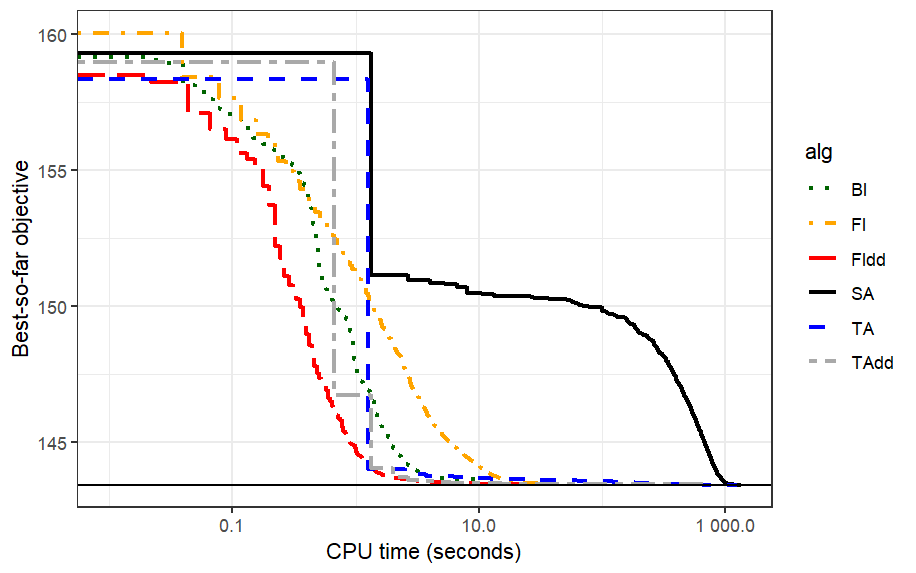}
        \caption{Best-so-far objective by logarithm time for Case III}
        \label{fig:bsf_30_40_8}

\end{figure}

\begin{figure}[h]
    \centering

        \includegraphics[scale=0.8]{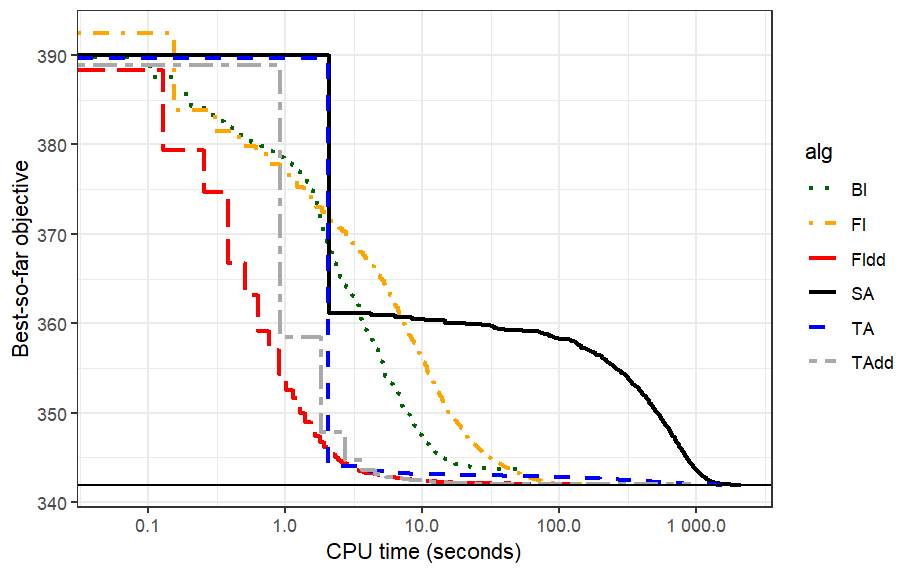}
        \caption{Best-so-far objective by logarithm time for Case IV}
        \label{fig:bsf_60_40_10}

\end{figure}

The Figures \ref{fig:bsf_5_40_2} to \ref{fig:bsf_60_40_10} show the best-so-far values vs. logarithmic time. The black horizontal line across all graphs is the best value found by any algorithm. This value is almost always found by the TA or SA algorithm. Both the TA and SA algorithms approach this value; however, they often do so in a substantially longer time than the other algorithms. The BI algorithm reaches values close to the best value found fast in Cases I and II. The FIdd and TAdd algorithms reach values close to the best value found faster than the other algorithms for Cases III and IV. After ten seconds, the FIdd and TAdd algorithms have reached the lowest value, even though SA and TA have a lower final criterion value. If calculating the relative D-efficiency for the final criterion value of FIdd or TAdd compared to, e.g., the final criterion value of SA, it can be seen that the number of individuals needed to get the same precision as SA is on the order of 1 \%.  

\clearpage

\clearpage

\section{Discussion}\label{Disc}

We considered a model-based approach for incomplete block designs in which block assignment depends on individual-specific covariates. We suggest and compare several algorithms for finding the optimal value $\Psi = -\sum_{i=1}^n\text{log[det}( M_i)]$ (D-optimality) where $M_i$ is the Fisher information matrix. 

We implemented existing local search methods such as simulated annealing, threshold accepting, and first- and best-improvement algorithms. We further utilized our model-based assumption and developed a first-improvement algorithm based on directional derivatives to be able to make wiser exchanges, which decreases the number of iterations and increases the speed. Since local search algorithms often end up in some local optimum, not close enough to the global optimum, we used a chained local optimization algorithm where we used the threshold accepting criterion and the first improvement algorithm based on directional derivatives as local search.

We tested our algorithms on a model-based optimal design problem in educational measurement called item calibration. The problem is to allocate examinees to items on an achievement test based on their estimated abilities, where the aim is to estimate the item parameters as efficiently as possible and not burden the examinees too much.

Testing the different algorithms on the item calibration problem in four different settings shows that the simulated annealing and the threshold algorithm find equal solutions in every case, but simulated annealing is more computationally intensive. 

The best improvement algorithm is really fast but gets stuck in a local optimum earlier than the other algorithms, which has less practical significance in the smaller problems. We can see that our idea to use directional derivatives makes the algorithms make wiser choices to approach the optimum faster, especially for the larger cases. Even though both the simulated annealing and the threshold algorithms find lower criterion values, it is of little practical meaning in terms of the number of individuals to achieve the same precision in the estimates of the item parameters.

A conclusion is that the simulated annealing or the threshold acceptance algorithms are good choices if time is not an issue. If we need faster algorithms, utilizing the directional derivatives is a smart strategy. We lose only a small amount of precision but gain a lot of computational time. This can be important, e.g., when the optimal design is applied adaptively. A disadvantage of several of the algorithms is that they require optimal hyperparameter values. The first improvement algorithm that uses directional derivatives is really fast and requires no hyperparameter decisions, which makes it a good option.

Future research can explore methods for finding good hyperparameters and also testing the algorithms on other problems.

\setcounter{secnumdepth}{0}

\clearpage

\bibliography{ref.bib}

@book{senn2002cross,
  title={Cross-over trials in clinical research},
  author={Senn, Stephen S},
  year={2002},
  publisher={John Wiley \& Sons}
}

@Manual{R-blocksdesign,
  title = {blocksdesign: Nested and Crossed Block Designs for Factorial and Unstructured Treatment Sets},
  author = {R. N. Edmondson},
  year = {2021},
  note = {R package version 4.9},
  url = {https://CRAN.R-project.org/package=blocksdesign},
}

@Manual{R-ibd,
  title = {ibd: Incomplete Block Design},
  author = {B N Mandal},
  year = {2024},
  note = {R package version 1.6},
  url = {https://CRAN.R-project.org/package=ibd},
}

@Manual{R-crossdes,
  title = {crossdes: Construction of Crossover Designs},
  author = {Martin Oliver Sailer},
  year = {2022},
  note = {R package version 1.1-2},
  url = {https://CRAN.R-project.org/package=crossdes},
}

@Inbook{SalinasRuiz2024,
author="Salinas Ru{\'i}z, Josafhat
and Montesinos L{\'o}pez, Osval Antonio
and Crossa, Jose",
title="Incomplete Block Design",
bookTitle="Introduction to Experimental Designs with {PROC GLIMMIX} of {SAS}: Applications in Food Science and Agricultural Science",
year="2024",
publisher="Springer Nature Switzerland",
address="Cham",
pages="99--129",
isbn="978-3-031-65575-3",
doi="10.1007/978-3-031-65575-3_5",
url="https://doi.org/10.1007/978-3-031-65575-3_5"
}

@book{dey2010incomplete,
  title={Incomplete block designs},
  author={Dey, Aloke},
  year={2010},
  publisher={World Scientific}
}

@incollection{nguyen2025incomplete,
  title={Incomplete block designs},
  author={Nguyen, Nam-Ky and Blagoeva, Kalina Trenevska},
  booktitle={International Encyclopedia of Statistical Science},
  pages={1173--1176},
  year={2025},
  publisher={Springer}
}

@book{stone2020sensory,
  title={Sensory evaluation practices},
  author={Stone, Herbert and Bleibaum, Rebecca N and Thomas, Heather A},
  year={2020},
  publisher={Academic press}
}

@article{Miller_Fackle-Fornius_2024, title={Parallel Optimal Calibration of Mixed-Format Items for Achievement Tests}, volume={89}, DOI={10.1007/s11336-024-09968-3}, number={3}, journal={Psychometrika}, author={Miller, Frank and Fackle-Fornius, Ellinor}, year={2024}, pages={903–928}}

@book{Baker2004,
address = {Boca Raton, Florida},
author = {Baker, Frank B. and Kim, Seock-Ho},
editor = {},
isbn = {9780824758257},
pages = {495},
publisher = {CRC Press},
title = {Item Response Theory: Parameter Estimation Techniques. 2nd Edition},
year = {2004}
}

@article{CookNachtsheim1980,
 ISSN = {00401706},
 URL = {http://www.jstor.org/stable/1268315},
 author = {R. Dennis Cook and Christopher J. Nachtsheim},
 journal = {Technometrics},
 number = {3},
 pages = {315--324},
 publisher = {[Taylor & Francis, Ltd., American Statistical Association, American Society for Quality]},
 title = {A Comparison of Algorithms for Constructing Exact {D}-Optimal Designs},
 urldate = {2025-02-28},
 volume = {22},
 year = {1980}
}

@article{CruzDorea1998,
 ISSN = {00219002},
 URL = {http://www.jstor.org/stable/3215524},
 author = {J. R. Cruz and C. C. Y. Dorea},
 journal = {Journal of Applied Probability},
 number = {4},
 pages = {885--892},
 publisher = {Applied Probability Trust},
 title = {Simple Conditions for the Convergence of Simulated Annealing Type Algorithms},
 urldate = {2025-02-28},
 volume = {35},
 year = {1998}
}

@article{Mitra-etal1986,
 ISSN = {00018678},
 URL = {http://www.jstor.org/stable/1427186},
 author = {Debasis Mitra and Fabio Romeo and Alberto Sangiovanni-Vincentelli},
 journal = {Advances in Applied Probability},
 number = {3},
 pages = {747--771},
 publisher = {Applied Probability Trust},
 title = {Convergence and Finite-Time Behavior of Simulated Annealing},
 urldate = {2025-02-28},
 volume = {18},
 year = {1986}
}

@book{Atkinsson,
address = {Oxford},
author = {Atkinsson, Anthony C and Donev, Alexander N and Tobias,  Randall D},
isbn = {9780199296590},
pages = {511},
publisher = {Oxford University Press},
title = {{Optimum experimental designs, with SAS}},
year = {2007}
}

@article{Kirkpatrick,
 author = {S. Kirkpatrick and JR. C. Gelat and M. Vecchi},
 journal = {Journal of Applied Probability},
 number = {4},
 pages = {885--895},
 publisher = {Applied Probability Trust},
 title = {Convergence Theorems for a Class of Simulated Annealing Algorithms on Rd},
 urldate = {2024-05-20},
 volume = {29},
 year = {1992}
}

@book{Reeves,
  editor = {Reeves, Colin R.},
  title = {Modern heuristic techniques for combinatorial problems},
  year = {1993},
  isbn = {0470220791},
  publisher = {John Wiley \& Sons, Inc.},
  address = {USA}
}

@article{KirkpatrickGelattVecchi,
 author = {S. Kirkpatrick and JR. C. Gelat and M. Vecchi},
 journal = {Science},
 number = {4598},
 pages = {671--680},
 publisher = {American Association for the Advancement of Science.},
 title = {Optimization by Simulated Annealing},
 volume = {220},
 year = {1983}
}

@book{Fedorov,
  author = {Fedorov, V.V.},
  title = {Theory of Optimal Experiments},
  year = {1972},
  isbn = {9780123942456},
  publisher = {Academic Press},
  address = {New York}
}

@book{Talbi,
  author = {Talbi, El-Ghazali},
  title = {Metaheuristics, from design to implementation},
  year = {2009},
  isbn = {978047027858},
  publisher = {John Wiley \& Sons, Inc.},
  address = {Hoboken, New Jersey}
}

@article{Hassan2019,
author = {Ul Hassan, Mahmood and Miller, Frank},
journal = {Psychometrika},
volume = {84},
pages = {1101-1128},
title = {Optimal Item Calibration for Computerized Achievement Tests},
year = {2019},
}

@incollection{Lourenco2003,  
    title     = {Iterated local search},              
    author    = {Lourenço, H. R. and Martin, O. C. and Stützle, T},                       
    pages     = {320--353},                                
    crossref  = {HandbookMetaheuristics2003}                     
}

@book{HandbookMetaheuristics2003,  
    year      = {2003},                              
    editor    = {F. Glover and G. Kochenberger},
    title     = {Handbook of Metaheuristics}, 
    booktitle = {Handbook of Metaheuristics},           
    publisher = {Kluwer Academic Publishers},              
    address   = {New York, Boston, Dordrecht, London, Moscow}                  
}

@article{ULHASSAN2021107177,
title = {An exchange algorithm for optimal calibration of items in computerized achievement tests},
journal = {Computational Statistics and Data Analysis},
volume = {157},
pages = {107-177},
year = {2021},
issn = {0167-9473},
author = {Ul Hassan, Mahmood and Miller, Frank}
}

@Inbook{Taillard2023,
author="Taillard, {\'E}ric D.",
title="A Short List of Combinatorial Optimization Problems",
bookTitle="Design of Heuristic Algorithms for Hard Optimization: With Python Codes for the Travelling Salesman Problem",
year="2023",
publisher="Springer International Publishing",
address="Cham",
pages="31--61",
isbn="978-3-031-13714-3",
doi="10.1007/978-3-031-13714-3_2",
url="https://doi.org/10.1007/978-3-031-13714-3_2"
}

@article{Bianchi2009Survey,
  author    = {Leonora Bianchi and Marco Dorigo and Luca Maria Gambardella and Walter J. Gutjahr},
  title     = {A Survey on Metaheuristics for Stochastic Combinatorial Optimization},
  journal   = {Natural Computing},
  volume    = {8},
  number    = {2},
  pages     = {239--287},
  year      = {2009},
  doi       = {10.1007/s11047-008-9098-4},
  url       = {https://doi.org/10.1007/s11047-008-9098-4}
}

@article{Baghel2012,
author = {Baghel, Malti and Agrawal, Shikha and Silakari, Sanjay},
title = { Survey of Metaheuristic Algorithms for Combinatorial Optimization },
journal = { International Journal of Computer Applications },
issue_date = { November 2012 },
volume = { 58 },
number = { 19 },
month = { November },
year = { 2012 },
issn = { 0975-8887 },
pages = { 21-31 },
numpages = {9},
url = { https://ijcaonline.org/archives/volume58/number19/9391-3813/ },
doi = { 10.5120/9391-3813 },
publisher = {Foundation of Computer Science (FCS), NY, USA},
address = {New York, USA}
}

@book{Silvey,
address = {London},
author = {Silvey, Samuel D},
isbn = {9789400959149},
pages = {86},
publisher = {Chapman and Hall},
title = {{Optimal design. Monographs on applied probability and statistics} (1st ed., Vol. 1)},
year = {1980}
}

@article{ATKINSON201481,
title = {Elemental information matrices and optimal experimental design for generalized regression models},
journal = {Journal of Statistical Planning and Inference},
volume = {144},
pages = {81-91},
year = {2014},
note = {International Conference on Design of Experiments},
issn = {0378-3758},
doi = {https://doi.org/10.1016/j.jspi.2012.09.012},
url = {https://www.sciencedirect.com/science/article/pii/S0378375812003060},
author = {Anthony C. Atkinson and Valerii V. Fedorov and Agnes M. Herzberg and Rongmei Zhang}

}

@book{SeberWild,
address = {Hoboken, New Jersey},
author = {G. A. F. Seber and C. J. Wild},
isbn = {9780471617600},
pages = {768},
publisher = {John Wiley and Sons, Inc.},
title = {Nonlinear Regression},
year = {1989}
}

@book{FedorovLeonov,
address = {Boca Raton},
author = {Fedorov, V.V. and Leonov, S.L.},
isbn = {9780429103988},
pages = {402},
publisher = {CRC Press},
title = {Optimal Design for Nonlinear Response Models (1st ed.) },
year = {2013}
}

@article{BiedermannWoods,
 ISSN = {00359254, 14679876},
 URL = {http://www.jstor.org/stable/41057575},
 author = {Stefanie Biedermann and David C. Woods},
 journal = {Journal of the Royal Statistical Society. Series C (Applied Statistics)},
 number = {2},
 pages = {281--299},
 publisher = {[Wiley, Royal Statistical Society]},
 title = {Optimal designs for generalized non-linear models with application to second-harmonic generation experiments},
 urldate = {2025-06-26},
 volume = {60},
 year = {2011}
}

@book{Papadimitriou1998,
address = {Mineola, N.Y.},
author = {Papadimitriou, Christos H. and Steiglitz, Kenneth},
isbn = {0486402584},
pages = {496},
publisher = {Dover},
title = {Combinatorial optimization: algorithms and complexity },
year = {1998}
}

@article{Bellman1962,
author = {Bellman, Richard},
title = {Dynamic Programming Treatment of the Travelling Salesman Problem},
year = {1962},
publisher = {Association for Computing Machinery},
address = {New York, NY, USA},
volume = {9},
number = {1},
issn = {0004-5411},
url = {https://doi.org/10.1145/321105.321111},
doi = {10.1145/321105.321111},
journal = {J. ACM},
pages = {61–63},
numpages = {3}
}

@book{Nemhauser1999,
author = {Nemhauser, George and Wolsey, Laurence},
isbn = {9780471828198},
pages = {783},
publisher = {John Wiley and Sons, Inc.},
title = {Integer and Combinatorial Optimization},
year = {1999}
}

@article{LandDoig1960,
 author = {A. H. Land and A. G. Doig},
 journal = {Econometrica},
 number = {3},
 pages = {497--520},
 publisher = {[Wiley, Econometric Society]},
 title = {An Automatic Method of Solving Discrete Programming Problems},
 volume = {28},
 year = {1960}
}

@article{Kong2021,
 author = {Kong, Yunfeng},
 journal = {Computational Urban Science},
 number = {19},
 publisher = {Springer},
 title = {An iterative local search based hybrid algorithm for the service area problem},
 volume = {1},
 year = {2021}
}

@book{AartsLenstra2003,
  editor    = {Aarts, Emile and Lenstra, Jan Karel},
  title     = {Local Search in Combinatorial Optimization},
  year      = {2003},
  publisher = {Princeton University Press},
  address   = {Princeton, NJ},
  note      = {Reprint. Originally published: New York : Wiley, 1997.},
  isbn      = {0-691-11522-2}
}

@article{Hamming1950,
  author    = {Hamming, Richard W.},
  title     = {Error Detecting and Error Correcting Codes},
  journal   = {Bell System Technical Journal},
  volume    = {29},
  number    = {2},
  pages     = {147--160},
  year      = {1950},
  doi       = {10.1002/j.1538-7305.1950.tb00463.x}
}

@book{Diestel2021,
  author    = {Diestel, Reinhard},
  title     = {Graph Theory},
  edition   = {2},
  year      = {2000},
  publisher = {Springer},
  address   = {New York},
  isbn      = {0-387-95014-1}
}

@article{GARCIARODENAS2020106844,
title = {A comparison of general-purpose optimization algorithms for finding optimal approximate experimental designs},
journal = {Computational Statistics and Data Analysis},
volume = {144},
pages = {106844},
year = {2020},
issn = {0167-9473},
doi = {https://doi.org/10.1016/j.csda.2019.106844},
url = {https://www.sciencedirect.com/science/article/pii/S0167947319301999},
author = {Ricardo García-Ródenas and José Carlos García-García and Jesús López-Fidalgo and José Ángel Martín-Baos and Weng Kee Wong}

}

@misc{alves2025unifiedbetaregressionmodel,
      title={Unified Beta Regression Model with Random Effects for the Analysis of Sensory Attributes}, 
      author={João César Reis Alves and Gabriel Rodrigues Palma and Idemauro Antonio Rodrigues de Lara},
      year={2025},
      eprint={2504.05996},
      archivePrefix={arXiv},
      primaryClass={stat.ME},
      url={https://arxiv.org/abs/2504.05996}, 
}

@article{DUECK1990161,
title = {Threshold accepting: A general purpose optimization algorithm appearing superior to simulated annealing},
journal = {Journal of Computational Physics},
volume = {90},
number = {1},
pages = {161-175},
year = {1990},
issn = {0021-9991},
doi = {https://doi.org/10.1016/0021-9991(90)90201-B},
url = {https://www.sciencedirect.com/science/article/pii/002199919090201B},
author = {Gunter Dueck and Tobias Scheuer}
}

@article{Martin1996,
title = {Combining simulated annealing with local search heuristics},
journal = {Annals of Operations Research},
volume = {63},
number = {1},
pages = {57-75},
year = {1996},
issn = {1572-9338},
doi = {https://doi.org/10.1007/BF02601639},
author = {Olivier C. Martin and Steve W. Otto}

}

\appendix
\section{}\label{appendix}

\begin{algorithm}[h]

\DontPrintSemicolon
\SetKwInOut{Input}{Input}
\SetKwInOut{Output}{Output}

\Input{Random starting solution $\mathbf{D}_{start}$ and associated information matrix $M_{start}$, initial temperature/threshold $t_0$, cooling factor $\alpha$, number of epochs $k$, number of repetitions per epoch $n_{rep}$}
\Output{Approximation $\mathbf{D}_{out}$ of the optimal solution}

Let $\mathbf{D}_{current} = \mathbf{D}_{start}$ and 
$M_{current} = M_{start}$

\For{j from 1 to k}{

    \For{i from 1 to $n_{rep}$}{
         $v = \delta(\mathbf{D}_{current}) \sim \text{Uniform}(\{1,\dots, b=|\mathcal{B}|\})$ where $\mathcal{B}$ is the set of columns in  $\mathbf{D}_{current}$ and apply $\phi_1(\mathbf{d}_v) =\mathbf{d}_v^{'}$ to get $\mathbf{D}_{cand}$. Compute $\Psi(M_{cand})$.
        
        \If{a set of columns $\mathcal{B^{'}}$ in $\mathbf{D}_{current}$ with 0 and 1 on opposite position compared to $v$ exists}{
            $v^{'} \sim \text{Uniform}(1,\dots,  |\mathcal{B}^{'}|)$, and apply $\tilde\phi_1(\mathbf{d}_v) =\mathbf{d}_{v^{'}}^{'}$ to get  $\mathbf{D}_{cand_2}$.
            Compute $\Psi(M_{cand_2})$. 
            
            \If{$\Psi(M_{cand2})$ < $\Psi(M_{cand})$}{
                $\mathbf{D}_{cand} = \mathbf{D}_{cand_2}$ 
            }
        }
        Let $\Delta E = \Psi(M_{cand}) - \Psi(M_{current})$.

        \If{(SA) $h > u$, where $u \sim \text{Uniform}[1,0]$ and   $h = \exp\left(\frac{\Delta E}{t_{j-1}}\right)$}{
             $\mathbf{D}_{current} = \mathbf{D}_{cand}$
        }
        \If{(TA) $\Delta E< t_{j-1}$}{
             $\mathbf{D}_{current} = \mathbf{D}_{cand}$
        }
    }
    
    Let $t_j = \alpha\cdot(t_{j-1})$
}
\Return{$\mathbf{D}_{out} = \mathbf{D}_{current}$}
\caption{Simulated annealing (SA)/Threshold accepting (TA) for minimizing the objective function $\Psi$ with solution $\mathbf{D}$ and associated information matrix $M$, 
inner iteration step $i$, and outer iteration step $j$}
\label{Alg:SATA}
\end{algorithm}

\begin{algorithm}[h]
\DontPrintSemicolon
\SetKwInOut{Input}{Input}
\SetKwInOut{Output}{Output}

\Input{Random starting solution $\mathbf{D}_{start}$ and associated information matrix $M_{start}$, number of blocks $b$, number of 0's $=n-d$ in each block,  number of 1's $=d$ in each block}
\Output{Approximation $\mathbf{D}_{out}$ of the optimal solution}

Let $\mathbf{D}_{current} = \mathbf{D}_{start}$,  
$M_{current} = M_{start}$, \texttt{improved} $\leftarrow$ \texttt{TRUE}, and $n_{exchange}=0$ \
 
 \While{\texttt{improved} == \texttt{TRUE}}{
   \texttt{improved} $\leftarrow$ \texttt{FALSE}
    
\For{i from 1 to b}{
$\delta(\mathbf{D}_{current}) = i$

    \For{j from 1 to d}{
       \For{k from 1 to $n-d$}{
       Define $ \phi_1(\mathbf{d}_i)=\mathbf{d}_i'$ by exchanging the j:th 1 to the k:th 0 to get $\mathbf{D}_{cand_{j,k}}$ and $M_{cand_{j,k}}$. Compute $\Psi(M_{cand_{j,k}})$. 
      
       }
       }
 For the $j$ and $k$ that minimizes $\Psi(M_{cand_{j,k}})$ let $M_{cand} =M_{cand_{j,k}}$ and $\mathbf{D}_{cand}=\mathbf{D}_{cand_{j,k}}$
       
        \If{a set of columns $\mathcal{B^{'}}$ in $\mathbf{D}_{current}$ with 0 and 1 on opposite position compared to $i$ exists}{
             $v^{'} \sim \text{Uniform}(1,\dots,  |\mathcal{B}^{'}|)$, and apply $\tilde\phi_1(\mathbf{d}_{v^{'}}) =\mathbf{d}_{v^{'}}^{'}$ to get $\mathbf{D}_{cand_2}$.
            Compute $\Psi(M_{cand_2})$.  

            \If{ $\Psi(M_{cand_2})$ <  $\Psi(M_{cand})$}{$ \mathbf{D}_{cand} = \mathbf{D}_{cand_2}$ }
            
            \If{$\Psi(M_{cand})$ <  $\Psi(M_{current})$}{$ \mathbf{D}_{current} = \mathbf{D}_{cand}$ and \texttt{improved} $\leftarrow$ \texttt{TRUE}.}
    
    }
    }
}
\Return{$\mathbf{D}_{out} = \mathbf{D}_{current}$}
\caption{Best improvement (BI) exchange algorithm for minimizing the objective function $\Psi$ with solution $\mathbf{D}$ and associated information matrix $M$, 
over version iteration step $i$, iteration step $j$ over 1's, iteration step $k$ over 0's.}
\label{Alg:BI}
\end{algorithm}

\begin{algorithm}[h]
\DontPrintSemicolon
\SetKwInOut{Input}{Input}
\SetKwInOut{Output}{Output}

\Input{Random starting solution $\mathbf{D}_{start}$ and associated information matrix $M_{start}$, number of blocks $b$, number of 0's $=n-d$ in each block,  number of 1's $=d$ in each block}
\Output{Approximation $\mathbf{D}_{out}$ of the optimal solution}

Let $\mathbf{D}_{current} = \mathbf{D}_{start}$,  
$M_{current} = M_{start}$, \texttt{improved} $\leftarrow$ \texttt{TRUE}, and $n_{exchange}=0$ \
 
 \While{\texttt{improved} == \texttt{TRUE}}{
   \texttt{improved} $\leftarrow$ \texttt{FALSE}
    
\For{i from 1 to b}{
$\delta(\mathbf{D}_{current}) = i$

For FIdd only: Calculate directional derivatives (dd) for all cells, and create a vector of differences of (dd) between all combinations of 1's and 0's, and sort  by biggest differences

\texttt{cond} $\leftarrow$ \texttt{FALSE}

    \For{j from 1 to d}{
    \If{\texttt{cond} $==$ \texttt{TRUE}}{break loop}
  
       \For{k from 1 to $n-d$}{
 Define $\phi_1(\mathbf{d}_i) = \mathbf{d}_i'$ by exchanging the j:th 1 to the k:th 0 (by dd difference for FIdd) to get $\mathbf{D}_{cand_{j,k}}$ and $M_{cand_{j,k}}$. Let $M_{cand} = M_{cand_{j,k}}$ and $\mathbf{D}_{cand} = \mathbf{D}_{cand_{j,k}}$

        \If{a set of columns $\mathcal{B^{'}}$ in $\mathbf{D}_{current}$ with 0 and 1 on opposite position compared to $i$ exists}{
             $v^{'} \sim \text{Uniform}(1,\dots,  |\mathcal{B}^{'}|)$, and apply $\tilde\phi_1(\mathbf{d}_{v^{'}}) =\mathbf{d}_{v^{'}}^{'}$ to get $\mathbf{D}_{cand_2}$.
            Compute $\Psi(M_{cand_2})$. 

            \If{ $\Psi(M_{cand_2})$ <  $\Psi(M_{cand})$}{$ \mathbf{D}_{cand} = \mathbf{D}_{cand_2}$ }
            
            \If{$\Psi(M_{cand})$ <  $\Psi(M_{current})$}{$ \mathbf{D}_{current} = \mathbf{D}_{cand}$, \texttt{improved} $\leftarrow$ \texttt{TRUE}, \texttt{cond} $\leftarrow$ \texttt{TRUE}, and break loop.}
    }
    }
       }
    }
    
}
\Return{$\mathbf{D}_{out} = \mathbf{D}_{current}$}
\caption{First improvement (FI)/First improvement with directional derivatives (FIdd) exchange algorithm for minimizing the objective function $\Psi$ with solution $\mathbf{D}$ and associated information matrix $M$, 
over block iteration step $i$, iteration step $j$ over 1's, iteration step $k$ over 0's}
\label{Alg:FIFIdd}
\end{algorithm}

\begin{algorithm}[h]
\DontPrintSemicolon
\SetKwInOut{Input}{Input}
\SetKwInOut{Output}{Output}

\Input{Random starting solution $\mathbf{D}_{start}$ and associated information matrix $M_{start}$, initial threshold $t_0$, number of epochs $k$, number of repetitions per epoch $n_{rep}$, perturb parameters $l$ and $n$, and number of $m$ consecutive updated $\mathbf{D}$ without improvement of $\Psi(M)$}
\Output{Approximation $\mathbf{D}_{out}$ of the optimal solution}

Let $\mathbf{D}_{current} = \mathbf{D}_{start}$,
$M_{current} = M_{start}$, and $s=0$

\For{j from 1 to k}{

    Let $\mathbf{D}_{global} =  \mathbf{D}_{current}$ and $M_{globas} = M_{current}$
    
    \For{i from 1 to $n_{rep}$}{
        
        perturb  $\mathbf{D}_{current}$ to get $\mathbf{D}_{pert}$ and run FIdd to get $\mathbf{D}_{cand}$ and $M_{cand}$
       
        Let $\Delta E = \Psi(M_{cand}) - \Psi(M_{current})$.

        \If{ $\Delta E< t_{j-1}$}{
             $\mathbf{D}_{current} = \mathbf{D}_{cand}$
        }

    }
      \If{$\Psi(M_{current}) > \Psi(M_{global})$}{
      s = s + 1}

  \Else{
    s=0
  }
  \If{s > m}{break loop}
  
    Let $t_j = t_{j-1}\cdot (1 - i/k)$
}
\Return{$\mathbf{D}_{out} = \mathbf{D}_{current}$}
\caption{Chained local optimization algorithm with FIdd as local search algorithm and threshold accepting criterion (TAdd) for minimizing the objective function $\Psi$ with solution $\mathbf{D}$ and associated information matrix $M$
}
\label{alg:TAdd}
\end{algorithm}

\end{document}